\documentclass[longbibliography,groupedaddress,showpacs,showkeys,amssymb,eqsecnum,aps,nofootinbib,superscriptaddress] {revtex4}
\usepackage[]{graphicx}
\usepackage[]{graphics}
\usepackage{amsmath}
\usepackage{epsf}                                                                                           
\usepackage{color}                   
\usepackage{verbatim}                                                                         
\usepackage{hyperref}

\newcommand{\be}{\begin{equation}}

\newcommand{\ee}{\end{equation}}

\begin{document}                                                                                              

\author{F. T. Brandt}
\email{fbrandtl@usp.br}
\affiliation{Instituto de F\'{\i}sica, Universidade de S\~ao Paulo, S\~ao Paulo, SP 05508-090, Brazil}

\author{J. Frenkel}
\email{jfrenkel@if.usp.br}
\affiliation{Instituto de F\'{\i}sica, Universidade de S\~ao Paulo, S\~ao Paulo, SP 05508-090, Brazil}

\author{S. Martins-Filho}   
\email{sergiomartinsfilho@usp.br}
\affiliation{Instituto de F\'{\i}sica, Universidade de S\~ao Paulo, S\~ao Paulo, SP 05508-090, Brazil}

\author{D. G. C. McKeon}
\email{dgmckeo2@uwo.ca}
\affiliation{
Department of Applied Mathematics, The University of Western Ontario, London, Ontario N6A 5B7, Canada}
\affiliation{Department of Mathematics and Computer Science, Algoma University, 
  Sault Ste.~Marie, Ontario P6A 2G4, Canada}

\author{G.  S.  S.  Sakoda}   
\email{gustavo.sakoda@usp.br}
\affiliation{Instituto de F\'{\i}sica, Universidade de S\~ao Paulo, S\~ao Paulo, SP 05508-090, Brazil}

\title{Forward scattering amplitudes in the imaginary time formalism}

\begin{abstract}
  We study, in the imaginary time formalism, the relation between loops and on-shell forward scattering tree amplitudes in thermal field theories. This allows for an efficient evaluation, at all temperatures,
of Green's functions with causal retarded boundary conditions. We present an application of this relation in quantum gravity coupled to scalar matter fields.
We show that at one or two loops, involving planar diagrams, the 1PI retarded thermal Green's functions may respectively be expressed in terms of connected forward scattering tree amplitudes of one or two on-shell particles.
\end{abstract}

\pacs{11.10.Wx,11.15.-q,04.60.-m}
\keywords{finite temperature field theory, gauge theories, gravity}

\maketitle

\section{Introduction}\label{sec1}
The connections between tree amplitudes and loop integrals play an important role in quantum field theory. In a pioneering work \cite{Feynman:1963ax},
Feynman has shown that one can relate one-loop integrals in gauge theories to on-shell forward tree scattering amplitudes. Such relations are useful due to the simplicity of these amplitudes, which may be obtained by contour integrations containing the poles of the propagators. There has been a lot of effort in the literature to find out how unitarity  methods, which rely on the fact that the loop amplitudes are determined by their singularities, can be used to evaluate higher order loops. These methods have been developed in a series of papers and successfully applied to loop computations in the standard model 
\cite{bern:1994zxbern:1995cg,Britto:2004nc,Anastasiou:2006gt,Forde:2007mi,Bern:2007dw}
When a single propagator in a one-loop graph is cut, the integrand leads to a tree-level
amplitude. By developing Feynman's method, some QCD amplitudes have been computed just from
single cuts. This approach has been extended, in the real time formulation at zero temperature, to higher-order loops, both for time-ordered \cite{Berger:2009zb,Brandhuber:2005kd,Rodrigo:2008fp,Bierenbaum:2010cy} and for retarded boundary conditions \cite{Caron-Huot:2010fvq}.
% 
%
%[7-11 ] ---> , both for time-ordered [ 7 - 10] as well as for 
%retarded [novo 11 = Caron-Huot ] boundary conditions .
%O que você acha ?

In the present work, we consider the retarded thermal Green's functions
that appear as causal response functions,
which are most conveniently studied in the framework of the imaginary time formalism. In this formulation, in momentum space, a quantum field theory in $3+1$ dimensions involves a 3-dimensional Euclidean theory with a summation over the discrete Matsubara frequencies $Q_0= 2\pi i n T$ where $T$ is the temperature \cite{kapusta:book89,lebellac:book96,das:book97}.
This allows one to separate the usual $T=0$ part from the finite temperature contribution. %One can then relate, in a unified and general way,
The main purpose of this work is to give a unified treatment of the relation between one- and two-loop graphs %to
and on-shell forward scattering tree amplitudes,
which is valid
both at zero as well as at finite temperature. This method, which greatly simplifies computations at finite temperature in thermal gauge theories,
has been previously used in connection with particular thermal amplitudes
\cite{Barton:1990fk,Frenkel:1992ts,Brandt:1993dk,Brandt:1999gf}.
Using appropriate analytic continuations of the external energies, one can recover the corresponding results obtained in the real time formalism.

In section \ref{sec2} we describe the forward scattering amplitude method, at one loop order, in the imaginary time formalism. %\cite{kapusta:book89,lebellac:book96}.
An application of this approach to the calculation of the lowest order retarded Green's functions in quantum gravity at zero temperature is given in section \ref{sec3}. The computations
are done in $D$ space-time dimensions using dimensional regularization 
\cite{tHooft:1972fi}. The extension of this analysis to finite temperature
is presented in section \ref{sec4}. A workable two loop example of this formulation in a scalar theory is examined in section \ref{sec5}. 
In section \ref{sec6}, we summarize the results and discuss a generalization of this approach at two-loop order. Some details of the calculations are given in the Appendices.

\section{The forward scattering method}\label{sec2}
\begin{figure}[b]
    \includegraphics[scale=0.5]{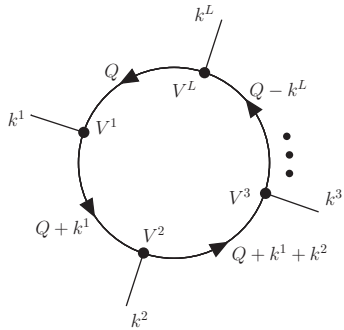}
    \caption{A generic one--loop diagram.}\label{fig2}
\end{figure}

In the imaginary time formalism, the one-loop integral associated with diagrams such as the one shown in Fig. \ref{fig2} is given by \cite{kapusta:book89,lebellac:book96}
\be
\mu^{4-D}
\sum_{\alpha,\,a,s}
\int\,\frac{{d}^{D-1} Q}{(2\pi)^{D-1}} \left[T \sum_{n=-\infty}^{\infty}
  f(Q_0,\vec Q,k_0^1,\vec k^1, \dots , k_0^L,\vec k^L)\right];\;\;\; 
(k^1+k^2+\dots+k^L=0)
\ee
where $D$ is the space-time dimension,
$Q_0 = i \pi (2 n + m)T$ are the {\it Matsubara frequencies}, and $m=0$ or
$m=1$ for Bosons or Fermions, respectively
($\sum_{\alpha,\,a,s}$
represents a sum over Lorentz, internal and spinor indices of the loop
particle).
The bosonic (fermionic) Matsubara summations can be done 
using the relation 
\begin{figure}[b]
\includegraphics[scale=0.4]{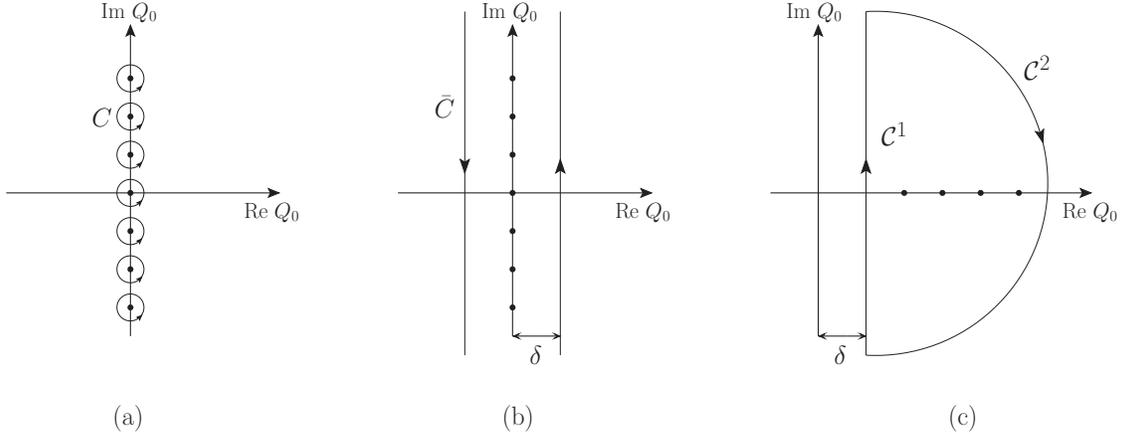}
\caption{Figures (a) and (b) show the equivalent integration contours in Eq. \eqref{e22}.
  Figure (c) shows the contour on the right hand side complex plane.
  Dots on the real $Q_0$ axis indicate the poles of Eq. \eqref{poles1}}\label{figa1}
\end{figure}
\be\label{e22}
  T \sum_{n=-\infty}^{\infty} f(Q_0)
  =
\oint_C \frac{d Q_0}{2\pi i} f(Q_0) \frac 1 2 \left[\coth\left(\frac{Q_0}{2T}\right)\right]^{\pm 1} 
  = \int_{-i\infty+\delta}^{i\infty+\delta} \frac{d Q_0}{2\pi i} 
\frac{f(Q_0)+f(-Q_0)}{2} 
\left(
1 \pm \frac{2}{e^{Q_0/T}\mp 1}
\right) 
\ee
(the equivalent contours are shown in figures (\ref{figa1}a) and (\ref{figa1}b))
which clearly allows us to separate the $T=0$ part from the finite-temperature contribution.  

Using Eq. \eqref{e22}, one-loop diagrams
containing $L$ internal lines, as shown in Fig. \ref{fig2},
can be expressed as
\begin{eqnarray}
  && \frac 1 2
     \sum_{\mu,\,a,s} 
     \int\,\frac{{d}^{D-1} Q}{(2\pi)^{D-1}}\int_{-i\infty+\delta}^{i\infty+\delta}
\frac{{d} Q_0}{2\pi i}\, \left[
{f(Q_0,\vec Q, k^1,k^2,\dots, k^L)+f(-Q_0,\vec Q, k^1,k^2,\dots, k^L)}\right]
\left(
1 \pm \frac{2}{e^{Q_0/T}\mp 1}
\right) 
     \nonumber \\ &=&
\frac 1 2
                 \sum_{\mu,\,a,s} 
                 \int\,\frac{{d}^{D-1} Q}{(2\pi)^{D-1}}\int_{-i\infty+\delta}^{i\infty+\delta}
\frac{{d} Q_0}{2\pi i}\, 
\left[f(Q_0,\vec Q, k^1,k^2,\dots, k^L)+ Q\leftrightarrow - Q
                      \right]
\left(
1 \pm \frac{2}{e^{Q_0/T}\mp 1}
\right) ,
\end{eqnarray}
where we have replaced $\vec Q$ by $-\vec Q$ in the second term of the integrand.
In general $f(Q_0,\vec Q, k^1,k^2,\dots, k^L)$ has the following structure
%(it is understood that the $i\eta$ terms are included in the denominators)
\be\label{poles1a}
f(Q_0,\vec Q, k^1,k^2,\dots, k^L) =
\frac{1}{{Q_0}^2-E^2(0,\vec Q)}\frac{1}{(Q_0+k^1_0)^2-E^2(\vec k^1,\vec Q)}
\cdots
\frac{t(Q, k^1,k^2\cdots k^L)}{(Q_0-k^L_0)^2-E^2(-\vec k^L,\vec Q)},
\ee
where $k^1+k^2+\cdots +k^L=0$, $E(\vec k^i,\vec Q)= \sqrt{|\vec Q+\vec k^i|^2 + m^2}$, 
and $t(Q, k^1,k^2\cdots k^L)$ is a tensor (or a scalar in the case
of a scalar field theory) which is determined by the interaction
vertices of the theory.  Using a partial fraction decomposition like
\be\label{poles1}
\frac{1}{(Q_0+{k}_0)^2-E^2(\vec k, \vec Q)} =
\frac{1}{2\,E(\vec k,\vec Q)}\left[
\frac{1}{Q_0+ k_0 - E(\vec k, \vec Q)} -
\frac{1}{Q_0+ k_0 + E(\vec k, \vec Q)}
\right]
\ee
($(k_0,\vec k)$ 
represents any of the following possibilities: $(0,\vec 0)$, $(k_0^1,\vec k^1)$, $(k_0^1+k_0^2,\vec k^1+\vec k^2)$, $\dots$, $(-k_0^L,-\vec k^L)$)
the $Q_0$ integral can be done upon
closing the integration contour on 
the right hand side of the complex plane
%(see Fig \ref{figcloseR})
and using Cauchy's integral formula
(assuming that
$f(Q_0,\vec Q, k^1,k^2,\dots, k^L)$ vanishes on the curve ${\cal C}^2$ in Fig.
\ref{figa1}c, when ${\cal C}^2$ is stretched to infinity).

Performing shifts in the loop momentum % $\vec Q$ 
\footnote{When these shifts occur in a linearly divergent integral in four
  dimensions, a surface term must also be included.
  This is needed to recover the usual axial anomaly \cite{Adler:1969gk}.}
%  (ref- Adler, S. L. (1969).  Physical Review. 177  2426–2438 )" 
and taking into account that the external frequencies
are integer multiple of $i 2\pi T$ so that 
$\exp(Q_0+i 2\pi \, l T)/T = \exp Q_0/T$ 
one can show that
%(in what follows we will restrict our analysis to the case of Bosonic fields)
\be\label{barton1}
%\left[
\begin{array}{l}
    \includegraphics[scale=0.5]{gen_one_loop.eps}
\end{array} %\right]_{T=0}
=  -  \sum_{\alpha,\,a,s} \mu^{4-D}
\int\,\frac{{d}^{D-1} Q}{(2\pi)^{D-1}} \frac{1}{2 E_Q}
\left(\frac 1 2 \pm \frac{1}{e^{E_Q/T}\mp 1}\right)
% \frac{1}{e^{|\vec Q|/T}-1}
%\left[ 
{\cal A}(k^1,k^2,\cdots,k^L;Q)
%+ Q\leftrightarrow  - Q
%\right]
,
\ee 
(the overall minus sign arises from the clockwise contour on the right hand
complex plane)
where $E_Q=E(0,Q)$ and
\begin{eqnarray}\label{genamp}
{\cal A}(k^1,k^2,\cdots,k^L;Q)&\equiv&
\left(
\begin{array}{c}
    \includegraphics[scale=0.7]{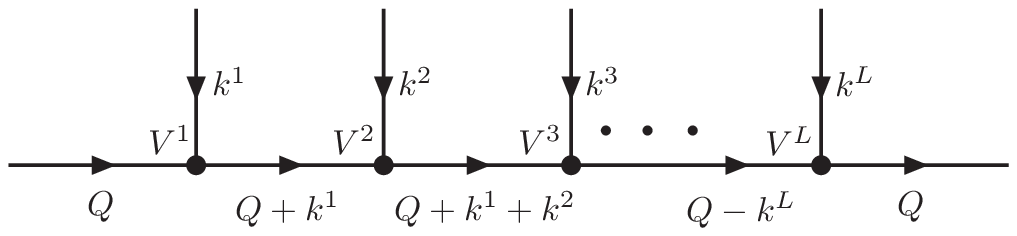}
\end{array}
+ Q\leftrightarrow  - Q
\right)_{Q_0=E_Q} 
\nonumber \\
&&\nonumber \\
&+& 
\mbox{ cyclic permutations of } V^1,V^2,\dots,V^L .
\end{eqnarray}
The tree amplitude ${\cal A}(k^1,k^2,\cdots,k^L;Q)$ is a forward 
scattering amplitude which describes the scattering of an
on-shell particle of momentum $Q$ by external particles of momentum
$k^1$, $k^2$, $\dots$, $k^L$.   %  \cite{Barton:1990fk}.
%It is
%a rational function of $k_0$
%which can be analytically continued to all continuous values of the
%external energy.

%\section{Examples}

%%%%%%%%%%%%%%%%%%%%%%%%%
\begin{comment}
\end{comment}
%%%%%%%%%%%%%%%%%%%%%%%%%%%%%%%
%%% \subsection{The self-energy in massless $g \phi^{3}$ theory}

The analytic continuation of some external energy $k_0^j\rightarrow k_0^j+ i \eta^j$, where $\eta^j$ is a positive infinitesimal quantity, with all other analytic continuations involving negative infinitesimals, yields the $j$-th retarded amplitude. In the following,
we shall not need to explicitly indicate this leg.

As a simple example, we consider
the  self-energy in massless $g \phi^{3}$ theory.
In this case, the forward scattering tree amplitude is
(where the on-shell condition $Q_0 = E(0,\vec Q)=|\vec Q|$ is to be understood)
\be\label{e28}
{\cal A}(k,Q) =  g^2
\left(\frac{1}{k^2+2 k \cdot Q} + \frac{1}{k^2-2 k \cdot Q} \right)
\ee
(we have taken into account the combinatorial factor $1/2$ and also added the permutation
as indicated in Eq. \eqref{genamp}), so that
the one-loop scalar self-energy can be written as
\begin{eqnarray}\label{2.3}
\Pi(k) =  -\frac{g^2}{2} \mu^{4-D}\int\,\frac{{d}^{D-1} Q}{(2\pi)^{D-1}}
  \frac{1}{2 |\vec Q|} 
\left(\frac{1}{k^2+2 k \cdot Q} + \frac{1}{k^2-2 k \cdot Q} \right).
\end{eqnarray}

%%%%%%%%%%%%%%%%%%%%%%%%%%%%%%%%%%%%%
\begin{comment}
\end{comment}
%%%%%
In the Appendix \ref{appA}, we compute the integral in \eqref{2.3} using the retarded analytic continuation of the external energy with $k_0\rightarrow k_0+i\eta$. This basic integral is also relevant for the self-energies of gauge fields and gravitons as we will see in the next section.

\begin{comment}
\end{comment}

\section{Scalar loops in a gravitational field}\label{sec3}
%\section{Scalar fields in curved space-time}\label{sec3}
The action for a scalar field in curved space-time is simply
\be\label{scalgravaction}
S = \frac{1}{2} \int d^D x \sqrt{-g}\left(
g^{\mu\nu} \partial_\mu\phi\partial_\nu\phi - m^2 \phi^2
\right).
\ee
For simplicity, let us set $m=0$. Defining
the gravitational field $\varphi^{\mu\nu}$
\be
\kappa\varphi^{\mu\nu} = \sqrt{-g} g^{\mu\nu} - \eta^{\mu\nu},
\ee
where $\kappa=\sqrt{16\pi G}$ is related to Newton's constant $G$,
the action becomes
\be\label{2.15}
S = \frac{1}{2} \int d^D x \left(-\phi\partial^2\phi +
\kappa\varphi^{\mu\nu} \partial_\mu\phi\partial_\nu\phi .
\right).
\ee
From \eqref{2.15} we obtain the following momentum space
Feynman rules: % (read from $i S$):
\begin{subequations}\label{scalFR}
\begin{eqnarray}
\begin{array}{l}
    \includegraphics[scale=0.7]{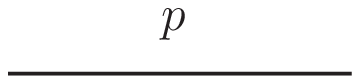}
\end{array}
  &:&  \;\;\;  \frac{i}{p^2},
%\ee
%\be
 \\
\begin{array}{l}
    \includegraphics[scale=0.7]{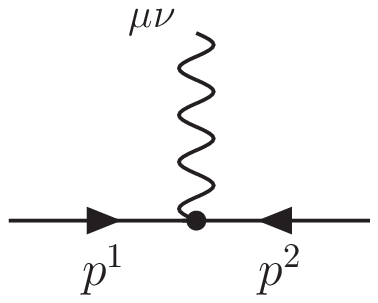}
\end{array} 
&:& \;\;\;  - \frac{i \kappa}{2}(p^1_\mu p^2_\nu +p^2_\mu p^1_\nu) ,
\end{eqnarray}
\end{subequations}        
where $p^1_\mu$ and $p^2_\nu$ are the momenta of the scalar field.

In terms of the forward scattering amplitude,
\begin{eqnarray}\label{genamp1}
  {\cal A}_{\mu\nu,\,\alpha\beta}(k,Q)
  & = &
\left(
\begin{array}{c}
    \includegraphics[scale=0.6]{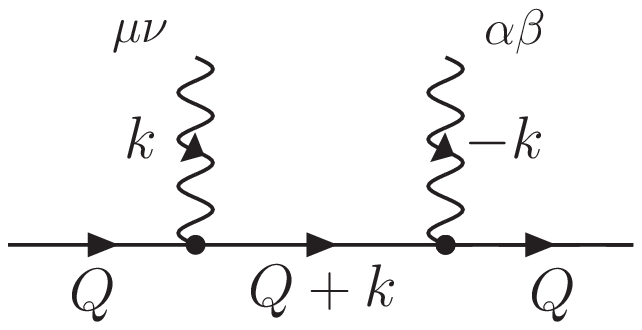}
\end{array}
+ Q\leftrightarrow  - Q
\right)_{Q_0=|\vec Q|} 
+ (\mu,\nu,k) \leftrightarrow (\alpha,\beta,-k)
\end{eqnarray}
the graviton self-energy at zero temperature can be written as
\be\label{2.17a}
\Pi_{\mu\nu,\,\alpha\beta}(k) = -\frac 1 2
\mu^{4-D}\int \frac{d^{D-1} Q}{(2\pi)^{D-1}} \frac{1}{2|\vec Q|}
{\cal A}_{\mu\nu,\,\alpha\beta}(k,Q),
\ee
where we have used the $T=0$ part of the general expression given in Eq. \eqref{barton1}.
Upon using the Feynman rules given in Eqs. \eqref{scalFR}, we obtain
\begin{eqnarray}\label{2.17}
{\cal A}_{\mu\nu,\,\alpha\beta}(k,Q) &=& \frac{\kappa^2}{4}
\left\{
  \frac{\left[Q_\mu(Q+k)_\nu + Q_\nu(Q+k)_\mu\right]\left[ 
    (Q+k)_\alpha Q_\beta+(Q+k)_\beta Q_\alpha\right]}
                                         {k^2+2k\cdot Q}
\right. \nonumber \\ 
                                         &+& \left .
  \frac{\left[Q_\alpha(Q-k)_\beta + Q_\beta(Q-k)_\alpha\right]\left[ 
    (Q-k)_\mu Q_\nu+(Q-k)_\nu Q_\mu\right]}
  {k^2-2k\cdot Q}
\right\} 
\end{eqnarray}
(we have included the combinatorial factor $1/2$ and also added
the permutation indicated in Eq. \eqref{genamp}).

After integration, \eqref{2.17} can be expressed
(by covariance) in terms of the five tensors shown 
in table I, so that 
\begin{equation}\label{2.19}
  \Pi_{\mu\nu,\, \alpha\beta}(k) =  \sum_{i=1}^{5}  {\cal T} ^i_{\mu\nu,\,  \alpha\beta}(k)
C_i(k) .
\end{equation}
The coefficients $C_i$ can be obtained solving
the following system of five algebraic equations
\begin{eqnarray}\label{2.22}
\sum_{i=1}^5 {\cal T}^i_{\mu\nu,\, \alpha\beta}(k) {\cal T}^j{}^{\mu\nu,\, \alpha\beta}(k) C_i(k) &=&
\Pi_{\mu\nu,\,\alpha\beta}(k) {\cal T}^j{}^{\mu\nu,\, \alpha\beta}(k) %\equiv
%J^j(k)
\nonumber \\ &=&
-\frac 1 2 \mu^{4-D}\int \frac{d^{D-1} Q}{(2\pi)^{D-1}} \frac{1}{2|\vec Q|}
{\cal A}_{\mu\nu,\, \alpha\beta}(k,Q)
                 {\cal T}^j{}^{\mu\nu,\, \alpha\beta}(k) ; \;\; j=1,\dots,5 .
\end{eqnarray}

\begin{table}[t]
\begin{center}
\begin{tabular}{c l}\hline \hline \\
${\cal T}^{1} _{\mu \nu ,\, \alpha \beta} (k)=$&$  \displaystyle{\frac{k_\mu k_\nu k_\alpha  k_\beta}{k^4}}$\\ & \\
${\cal T}^2 _{\mu \nu ,\, \alpha \beta} (k)=$&$ \displaystyle{\eta_{\mu \nu} \eta_{\alpha\beta}}$ \\ & \\
${\cal T}^{3}_{\mu \nu ,\,  \alpha \beta}(k)=$&$ \displaystyle{ \eta _{\mu \alpha} \eta_{\nu \beta} + \eta _{\mu \beta} \eta_{\nu \alpha}}$  \\ & \\
${\cal T}^4 _{\mu \nu ,\, \alpha \beta} (k)=$&$ \displaystyle{\frac{\eta_{\mu \nu} k_\alpha k_\beta + \eta_{\alpha \beta} k_\mu k_\nu}{k^2}}$ \\ & \\
${\cal T}^5 _{\mu \nu ,\,  \alpha \beta}(k)=$&$ \displaystyle{\frac{\eta_{\mu \alpha} k_\nu k_\beta +
\eta_{\mu \beta} k_\nu k_\alpha + \eta_{\nu \alpha} k_\mu k_\beta +
\eta_{\nu \beta} k_\mu k_\alpha}{k^2}}$ 
\\  \\ \hline \hline 
\end{tabular}\caption{The five independent tensors built from 
$\eta_{\mu\nu}$ and $k_\mu$, satisfying the symmetry conditions 
${\cal T}^{i}_{\mu\nu,\, \alpha\beta} (k)= {\cal T}^{i}_{\nu\mu,\, \alpha\beta}(k)= 
{\cal T}^{i}_{\mu\nu,\, \beta\alpha} (k)= {\cal T}^{i}_{\alpha\beta,\, \mu\nu}(k)$.}
\end{center}
\end{table}\label{tab1}

Using simple identities such as
\be
\frac{(k\cdot Q)^2}{k^4 - 4 (k\cdot Q)^2}
= \frac 1 4 \frac{k^4}{k^4 - 4 (k\cdot Q)^2} - \frac 1 4
\ee
and setting equal to zero any dimensional regularized integrals
that are independent of $|\vec Q|$, then all the integrals on the right-hand side
of Eq. \eqref{2.22} can be reduced to 
the following basic scalar integral 
\begin{eqnarray}\label{basicI}
I(k) &=& \kappa^2  \mu^{4-D}\int\,\frac{{d}^{D-1} Q}{(2\pi)^{D-1}}
  \frac{1}{2 |\vec Q|} 
  \left(\frac{1}{k^2+2 k \cdot Q} + \frac{1}{k^2-2 k \cdot Q} \right)
\nonumber \\
     &=&
-\kappa^2 \mu^{4-D} \frac{1}{(4\pi)^{D/2}} \Gamma(2-D/2)
\frac{\Gamma^2(D/2-1)}{\Gamma(D-2)} 
  \left[-i\eta\; {\rm sgn}(k_0) -k^2\right]^{D/2-2},
\end{eqnarray}
where we have used the result given in Eq. \eqref{A8} (up to an overall factor). 

Solving the linear system of equations in Eq. \eqref{2.22}, we obtain
\begin{comment} %%%%%%%%%%%%%%%%%%%%%%%%%%%%%%%%%%
\end{comment} %%%%%%%%%%%%%%%%%%%%%%%%%%%%%%%%%%%%
%(using Mathematica) 
\begin{subequations}\label{C1to5}
\be
C_1(k) = \frac{D(D-2)}{32 (D^2-1)} k^4 I(k),
\ee
\be
C_2(k) = \frac{k^4}{32(D^2-1)} I(k),
\ee
\be
C_3(k) = C_2(k) = - C_5(k) ,
\ee
and
\be
C_4(k) = \frac{D}{32(D^2-1)} k^4 I(k)
\ee
%and
%\be
%C_5(k) = -C_3(k).
%\ee
\end{subequations}
Substituting \eqref{C1to5} into \eqref{2.19}, we obtain
%The full tensor structure is given by
\begin{eqnarray}\label{2.25}
\Pi_{\mu\nu,\,\alpha\beta}(k) &=&\left[ 
\frac{D(D-2)k_\mu k_\nu k_\alpha k_\beta}{k^4} +
\eta_{\mu\nu} \eta_{\alpha\beta}+
\eta_{\mu\alpha} \eta_{\nu\beta}+
\eta_{\mu\beta} \eta_{\nu\alpha}  
\right. \nonumber \\ 
&+& \left.
\frac{D}{k^2} (\eta_{\mu\nu} k_\alpha k_\beta+\eta_{\alpha\beta} k_\mu k_\nu)
-\frac{\eta_{\mu \alpha} k_\nu k_\beta +
\eta_{\mu \beta} k_\nu k_\alpha + \eta_{\nu \alpha} k_\mu k_\beta +
\eta_{\nu \beta} k_\mu k_\alpha}{k^2}
\right] \frac{k^4I(k)}{32(D^2-1)}  .
\end{eqnarray}

This result %for the graviton self-energy in Eq. \eqref{2.25}
satisfies the Ward identity
\be\label{Ward0}
(\eta^{\mu\rho} k^\nu + \eta^{\nu\rho} k^\mu - \eta^{\mu\nu} k^{\rho})
  \Pi_{\mu\nu,\,\alpha\beta} = 0 ,
  \ee
which is stronger than the identity of the pure gravitational case in
the Eq. (3.36) of \cite{Brandt:2016eaj} (see also \cite{Lavrov:2019nuz}).
This can be understood since there is no BRST symmetry in the scalar case. There is only the classical diffeomorphism invariance of \eqref{scalgravaction} under
\begin{equation}\label{e17a}
  \delta \tilde g^{\mu\nu} = \kappa \delta (\phi^{\mu\nu})
  = \tilde g^{\mu\rho} \partial_\rho \theta^\nu +
                           \tilde g^{\nu\rho} \partial_\rho \theta^\mu
                          -\partial_\rho(\tilde g^{\mu\nu}\theta^\rho);
\;\;\; \tilde g^{\mu\nu} = \sqrt{-g} g^{\mu\nu}.
\end{equation}
associated with the coordinate transformations
$x^\mu\rightarrow x^\mu - \theta^\mu$.

\section{Graviton thermal self-energy}\label{sec4}
% \section{Finite Temperature}\label{sec4}
%%################################################################### xxxxxxxxx
Let us now consider the thermal part of Eq. \eqref{barton1} which contains the Bose-Einstein distribution.
When the external momenta $k$ are such that $k\ll T$ the loop integral
is dominated by the {\it hard thermal loop} region.
Using expansions like
\be
\frac{1}{k^2 + 2 k\cdot Q} = \frac{1}{2 k\cdot Q} 
- \frac{k^2}{(2 k\cdot Q) ^2} + \dots  ,
\ee
($k^2 \ll k\cdot Q$)
the tree amplitude in Eq. \eqref{2.17} reduces to
\begin{comment}
\end{comment}
\be\label{Ahtl1}
{\cal A}^{HTL}_{\mu\nu,\,\alpha\beta}(k,Q) = \frac{\kappa^2}{2}|\vec Q|^2
\left(
\frac{k_\mu \hat Q_\nu \hat Q_\alpha \hat Q_\beta}{k\cdot \hat Q}+
\frac{\hat Q_\mu k_\nu \hat Q_\alpha \hat Q_\beta}{k\cdot \hat Q}+
\frac{\hat Q_\mu \hat Q_\nu k_\alpha \hat Q_\beta}{k\cdot \hat Q}+
\frac{\hat Q_\mu \hat Q_\nu \hat Q_\alpha k_\beta}{k\cdot \hat Q}-
\frac{k^2 \hat Q_\mu \hat Q_\nu \hat Q_\alpha \hat Q_\beta}{(k\cdot \hat Q)^2}
\right)
\ee
to leading order in  the high temperature expansion ($\hat Q = Q/|\vec Q|$).

Inserting Eq. \eqref{Ahtl1} into the thermal part of 
Eq. \eqref{barton1} 
we obtain the following expression for
the thermal scalar loop contribution to the graviton self-energy 
\begin{eqnarray}
  \Pi^{HTL}_{\mu\nu,\,\alpha\beta}
\begin{comment}
\end{comment}
   &=&
-\frac{\kappa^2}{4} \frac{\mu^{4-D} T^{D}}{(2\pi)^{D-1}}
\int_0^{\infty} 
\frac{u^{D-1} du}{e^{u}-1}
    \nonumber \\                           &&
                               \int{d\Omega}
\left(
\frac{k_\mu \hat Q_\nu \hat Q_\alpha \hat Q_\beta}{k\cdot \hat Q}+
\frac{\hat Q_\mu k_\nu \hat Q_\alpha \hat Q_\beta}{k\cdot \hat Q}+
\frac{\hat Q_\mu \hat Q_\nu k_\alpha \hat Q_\beta}{k\cdot \hat Q}+
\frac{\hat Q_\mu \hat Q_\nu \hat Q_\alpha k_\beta}{k\cdot \hat Q}-
\frac{k^2 \hat Q_\mu \hat Q_\nu \hat Q_\alpha \hat Q_\beta}{(k\cdot \hat Q)^2}
    \right).
    \nonumber \\
\end{eqnarray}
Upon integrating over $u$ %Gradshteyn {\bf 3.411}-1
we obtain %\cite{gradshteyn} 
\begin{eqnarray}
  \Pi^{HTL}_{\mu\nu,\,\alpha\beta} &=&
-\frac{\kappa^2}{4} \frac{\mu^{4-D} T^{D}}{(2\pi)^{D-1}}
\Gamma(D)\zeta(D)
    \nonumber \\
&&    \int{d\Omega}
\left(
\frac{k_\mu \hat Q_\nu \hat Q_\alpha \hat Q_\beta}{k\cdot \hat Q}+
\frac{\hat Q_\mu k_\nu \hat Q_\alpha \hat Q_\beta}{k\cdot \hat Q}+
\frac{\hat Q_\mu \hat Q_\nu k_\alpha \hat Q_\beta}{k\cdot \hat Q}+
\frac{\hat Q_\mu \hat Q_\nu \hat Q_\alpha k_\beta}{k\cdot \hat Q}-
\frac{k^2 \hat Q_\mu \hat Q_\nu \hat Q_\alpha \hat Q_\beta}{(k\cdot \hat Q)^2}
    \right),
\end{eqnarray}
where $\Gamma$ and $\zeta$ denote respectively the Gamma and Zeta functions.
For $D=4$, we obtain 
\begin{eqnarray}
  \Pi^{HTL}_{\mu\nu,\,\alpha\beta} &=&
-\frac{\kappa^2 \pi^2 T^4}{120} 
%    \nonumber \\             &&
                               \int \frac{d\Omega}{4\pi}
\left(
\frac{k_\mu \hat Q_\nu \hat Q_\alpha \hat Q_\beta}{k\cdot \hat Q}+
\frac{\hat Q_\mu k_\nu \hat Q_\alpha \hat Q_\beta}{k\cdot \hat Q}+
\frac{\hat Q_\mu \hat Q_\nu k_\alpha \hat Q_\beta}{k\cdot \hat Q}+
\frac{\hat Q_\mu \hat Q_\nu \hat Q_\alpha k_\beta}{k\cdot \hat Q}-
\frac{k^2 \hat Q_\mu \hat Q_\nu \hat Q_\alpha \hat Q_\beta}{(k\cdot \hat Q)^2}
    \right)
    \nonumber \\
\end{eqnarray}
which is half the result in the case of pure gravity (A graviton loop produces a factor of two,
which is associated with the extra degree of freedom of a massless spin 2 particle).
%{\bf (there is an extra minus sign in relation to the 1993 papers)}.

\begin{comment}
\end{comment}

\begin{figure}[t]
    \includegraphics[scale=0.7]{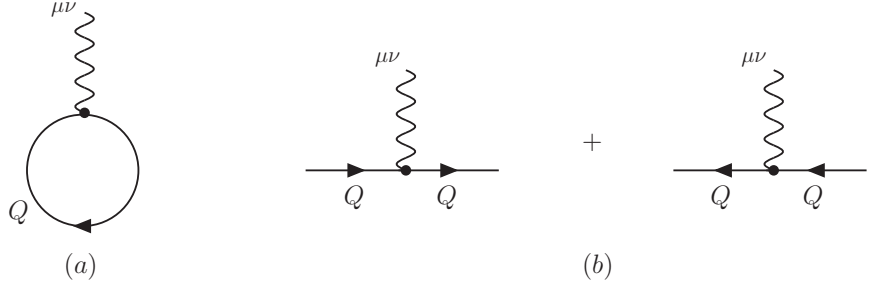}  
  \caption{The one-graviton thermal amplitude and the corresponding
    forward scattering amplitude.}\label{figtad}
\end{figure}

At finite temperature, the tadpole diagram shown in Fig. \ref{figtad}a does not vanish.
%(the corresponding forward scattering amplitude in Fig. \ref{figtad}b).
Taking into account the symmetry factor $1/2$ and the Feynman rules in \eqref{scalFR}
Fig. \ref{figtad}b yields
\be\label{tadAT}
{\cal A}_{\mu\nu}(k,Q) = -\kappa \, Q_\mu Q_\nu.
\ee
We can verify that
the amplitudes in Eqs. \eqref{Ahtl1} and \eqref{tadAT} are related
by the Ward identity
\be
\left(
\eta^{\mu\rho} k^\nu + \eta^{\nu\rho} k^\mu - \eta^{\mu\nu} k^{\rho}
\right)
\; {\cal A}^{HTL}_{\mu\nu,\,\alpha\beta}(k,Q) = -\kappa\,
\eta^{\nu\rho}\left(k_\alpha \delta^\mu_\beta + k_\beta \delta^\mu_\alpha\right)
\; {\cal A}_{\mu\nu}(Q). 
\ee
Since this identity is verified at the level of the tree amplitude,
the thermal self-energy and the one-graviton are also related by
the same identity, namely
\be
\left(\eta^{\mu\rho} k^\nu + \eta^{\nu\rho} k^\mu - \eta^{\mu\nu} k^{\rho}\right)
{\Pi }^{HTL}_{\mu\nu,\,\alpha\beta} = -\kappa\,
\eta^{\nu\rho}\left(k_\alpha \delta^\mu_\beta+k_\beta \delta^\mu_\alpha\right)
{\Pi}^{tadpole}_{\mu\nu},
\ee
in the hard thermal loop limit. Notice that at $T=0$ Eq. \eqref{Ward0}
%
%\be
%(\eta^{\mu\rho} k^\nu + \eta^{\nu\rho} k^\mu - \eta^{\mu\nu} k^{\rho})
%{\Pi }^{T=0}_{\mu\nu,\,\alpha\beta}(k,Q) = 0,
%\ee
is consistent with the absence of tadpoles.

In fact, it is straightforward to verify that the {\it exact amplitude} in 
Eq. \eqref{2.17} satisfy the same Ward identity as the hard thermal loop
case, namely
\be
\left(
\eta^{\mu\rho} k^\nu + \eta^{\nu\rho} k^\mu - \eta^{\mu\nu} k^{\rho}
\right)
\; {\cal A}_{\mu\nu,\,\alpha\beta}(k,Q) = -\kappa\,
\eta^{\nu\rho}\left(k_\alpha \delta^\mu_\beta + k_\beta \delta^\mu_\alpha\right)
\; {\cal A}_{\mu\nu}(Q),
\ee
At $T=0$,
\be
-\frac 1 2 \int\frac{d^{D-1} Q}{(2\pi)^{D-1}} \frac{1}{2 |\vec Q|}{\cal A}_{\mu\nu}(Q)=
\frac 1 4 \int\frac{d^{D-1} Q}{(2\pi)^{D-1}} 
\frac{Q_\mu Q_\nu}{|\vec Q|} = 0
\ee
and at finite $T$ we have 
\be 
- \mu^{4-D}\int\frac{d^{D-1} Q}{(2\pi)^{D-1}}\frac{1}{2|\vec Q|}\frac{{\cal A}_{\mu\nu}(Q)}
{e^{|\vec Q|/T} - 1}=
\mu^{4-D}\frac{\kappa}{2} \int\frac{d^{D-1} Q}{(2\pi)^{D-1}} \frac{1}{|\vec Q|}\frac{Q_\mu Q_\nu}
{e^{|\vec Q|/T} - 1} = C \mu^{4-D} T^{D}
\left(D u_\mu u_\nu - \eta_{\mu\nu}\right),
\ee
where $C$, is a constant, $u = (1,0,0, \dots ,0)$ is a rest frame four-velocity
and we have used that the result is traceless, since $Q^2=0$.
From the  $00$ component 
\be\label{A12}
\mu^{4-D}\frac{\kappa}{2} \int\frac{d^{D-1} Q}{(2\pi)^{D-1}} \frac{|\vec Q|}{e^{|\vec Q|/T} - 1}
= (D-1) C \mu^{4-D} T^{D},
\ee
we obtain \cite{gradshteyn}
\begin{comment}
\end{comment}
\be
C = \mu^{4-D} \frac{\kappa}{2 (D-1) (2 \pi)^{D-1}} \frac{2 \pi^{\frac{D-1}{2}}}
{\Gamma\left(\frac{D-1}{2}\right)}
\Gamma(D) \zeta(D)
= \frac{\mu^{4-D} \kappa}{2^{D-1} \pi^{\frac{D-1}{2}}}\frac{\Gamma(D-1)}
{\Gamma\left(\frac{D-1}{2}\right)}\zeta(D).
\ee
For $D=4$ we obtain,
\be
C = \frac{\kappa}{8 \pi \pi^{1/2} \Gamma(3/2)} \Gamma(3) \zeta(4)
%= \frac{\kappa}{2 \pi^2}  \frac{\pi^4}{90}
= \frac{\kappa\pi^2}{180},
%= \frac{\rho_S}{6}.
\ee
which is half the result obtained in pure gravity. %\cite{Rebhan:1990yr},
%{\bf with the same sign}
%(notice that the one-graviton function is half the energy-momentum tensor; also Rebhan considers
%$\delta \Gamma/\delta g^{\mu\nu} =\delta \Gamma/\delta \tilde g^{\mu\nu} $)
\begin{comment}
\end{comment}

\begin{comment}%%########################################
\end{comment}%%########################################

\section{Retarded amplitudes at two loops}\label{sec5}
In order to study higher-order thermal loop contributions in the imaginary time formalism,  we consider,  for simplicity the scalar $\phi^4$ field theory,  described by the Lagrangian
\begin{equation}\label{e51}
{\cal L} =  \frac 1 2 \partial_\mu \phi \partial^\mu \phi - \frac{m^2}{2} \phi^2 - \frac{\lambda}{4!} \phi^4
\end{equation}
and examine the  retarded self-energy of the scalar field which can be obtained by the prescription
$k_0 \rightarrow k_0 + i \eta$. The one-loop contribution can be expressed in terms of a forward scattering amplitude for a on-shell particle as discussed in Eqs. \eqref{barton1} and \eqref{genamp}.  

\begin{figure}[b]
    \includegraphics[scale=0.6]{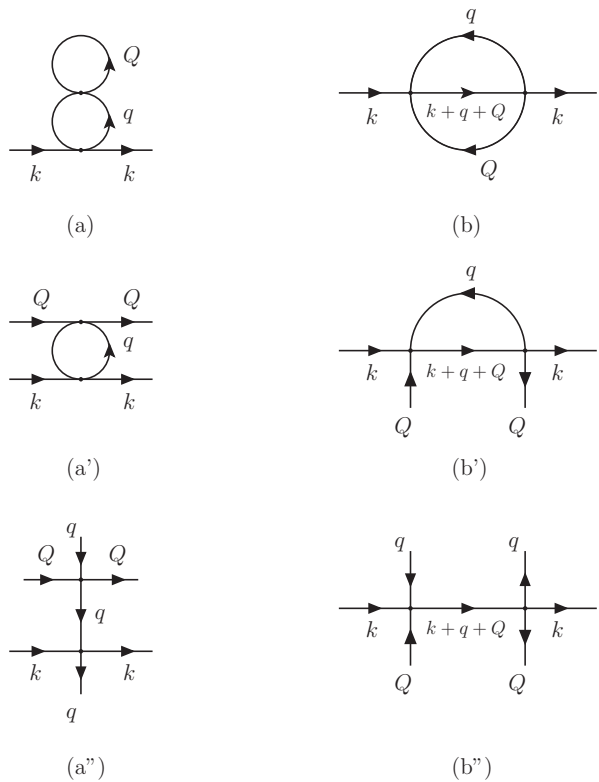}
    \caption{Two-loop
self-energy diagrams (a,b) and the corresponding forward amplitudes obtained by opening, successively, the $Q$ (a', b') and $q$ (a'', b'') internal lines.
    }\label{fig4}
\end{figure}

In going at two loops,  we note that there are two types of Feynman diagrams, as shown in the Figs.  (\ref{fig4}a) and (\ref{fig4}b).
%Evaluating first the $Q_0$ frequency sum by contour integration in the right hand side of the complex $Q_0$-plane (see Eq. \eqref{e22}), opens the $Q$-propagator and sets the $Q$-particle on-shell with a positive energy $Q_0=E_Q=\sqrt{|\vec Q|^2+m^2}$.   According to Eqs. \eqref{barton1} and \eqref{genamp},  this leads to a set of diagrams  of the form shown in Figs. (\ref{fig4}a') and (\ref{fig4}b' ),  where graphs obtained by permutations of the vertices as well as by the replacement $Q \rightarrow -Q$  are to be understood. Since at this stage $k_0=i2\pi l T$ and $q_0=i2\pi n T$, these graphs are multiplied by a phase-space factor which depends only on $Q_o=E_Q$. 
%%%%%%%%%%%%%%%%%%%%%%%%%%%%%%%%%%%%%%%%%%
\begin{comment}
\end{comment}
%%%%%%%%%%%%%%%%%%%%%%%%%%%%%%%%%%%%%%%%%%%%%%%%%%%%%
We can evaluate the summations over the discrete frequencies $Q_0 = i 2\pi n T$ and $q_0 = i 2 \pi s T$ with the help of the relation \eqref{e22}.  Next,  we proceed to evaluate the $Q_0$ and $q_0$ integrals by contour integrations in the right hand side of their complex planes.

In the case  of Fig.  (\ref{fig4}a), according to Eq. \eqref{barton1} this opens successively the Feynman propagators, and sets the $Q$ and $q$ particles on their mass-shell, with positive energies $Q_0$ = $E_Q$ and $q_0 = E_q$. In Fig. (\ref{fig4}a'), which has a double pole, one can do the $q_0$ integral by using Cauchy theorem, leading to a  graph like that shown in Fig. (\ref{fig4}a''). This diagram, which corresponds to the forward scattering amplitudes of two on-shell particles  is multiplied by a phase-space integral, leading to a k-independent result given by
\begin{equation}\label{e52}
\frac{\lambda^2}{4}
\int\frac{d^{D-1} Q}{(2\pi)^{D-1}}
\frac{1}{2 E_Q} \int\frac{d^{D-1} q}{2 E_q}
\left[\frac 1 2 +\frac{1}{e^{E_Q/T}-1}\right]
\left[\frac 1 2 +\frac{1}{e^{E_q/T}-1}\right]
\frac{-1}{|\vec q|^2+m^2}
\end{equation}
%On the other hand, the application of
%%%%%%%%%%%%%%%%%%%%%%%%%%%%%%%%%%%%%%%%%%%%%%%%%%%%%%%%%
\begin{comment}
\end{comment}
%%%%%%%%%%%%%%%%%%%%%%%%%%%%%%%%%%%%%%%%%%%%%%%%%%%%%%%%%%%%%%%%%%%%%%
%%%% Version of 20/08/2021 yyyyyyyyyyyyyyyyyyyyyyyyyyyyyyyyyyyyyyyyyy
We will next apply this method to the diagram shown in Fig. (\ref{fig4}b), which yields a $k$-dependent contribution given by 
\be\label{e53nv}
-\frac{\lambda^2}{3!}
\int\frac{d^{D-1} Q}{(2\pi)^{D-1}} \int\frac{d^{D-1} q}{(2\pi)^{D-1}}
T^2 \sum_{n,s}
\frac{1}{Q^2-m^2}\frac{1}{q^2-m^2}\frac{1}{(k+Q+q)^2-m^2},
\ee
where $Q_0=i 2\pi n T$, $q_0=i 2\pi s T$ and $k_0=i 2\pi l T$.
Performing the summations with the help of Eq. \eqref{e22} and replacing $\vec Q\rightarrow -\vec Q$ and $\vec q\rightarrow -\vec q$ in some terms,
we obtain for the integrand of \eqref{e53nv} the result 
\begin{eqnarray}\label{e54nv}
I &=& \int_{-i\infty}^{i\infty} \frac{dq_0}{2\pi i}\left[\frac{1}{2}+N(q_0)\right]
    \int_{-i\infty}^{i\infty} \frac{dQ_0}{2\pi i}\left[\frac{1}{2}+N(Q_0)\right]
\frac{1}{Q^2-m^2}\frac{1}{q^2-m^2}
\nonumber \\
& & \left[\frac{1}{(k+Q+q)^2-m^2}+
      \frac{1}{(k-Q-q)^2-m^2}+
      \frac{1}{(k+Q-q)^2-m^2}+
      \frac{1}{(k-Q+q)^2-m^2}
\right],
\end{eqnarray} 
where $N(\omega)$ denotes the statistical factor
\be\label{e55nv}
N(\omega) = \frac{1}{e^{\omega/T}-1}. 
\ee
We first perform the $Q_0$ integration along the contour shown in
Fig.(\ref{figa1}c) by using Cauchy's theorem. Making appropriate shifts and using the fact that $k_0$ is an integer multiple of $i2\pi T$ we obtain the following two contributions      
\begin{eqnarray}\label{e56nv}
I_1 &=& -\frac{1}{2 E_Q}
\int_{-i\infty}^{i\infty} \frac{dq_0}{2\pi i}\left[\frac{1}{2}+N(q_0)\right]
\frac{1}{q^2-m^2} \left[\frac{1}{2}+N(E_Q)\right]
\nonumber \\
&&
\left[\frac{1}{(k+Q+q)^2-m^2}+\frac{1}{(k-Q-q)^2-m^2}+
\frac{1}{(k+Q-q)^2-m^2}+\frac{1}{(k-Q+q)^2-m^2}
\right]_{Q_0=E_Q},
\end{eqnarray}
and
\begin{eqnarray}\label{e57nv}
I_2 &=& -\frac{1}{2 E_Q}
\int_{-i\infty}^{i\infty} \frac{dq_0}{2\pi i}\left[\frac{1}{2}+N(q_0)\right]\frac{1}{q^2-m^2} 
\nonumber \\ &&
\left\{\left[\frac{1}{2}+N(E_Q-q_0)\right]
\left[\frac{1}{(k+Q-q)^2-m^2}+\frac{1}{(k-Q+q)^2-m^2}\right]
\right. 
\nonumber\\ &+& \left. 
\left[\frac{1}{2}+N(E_Q+q_0)\right]
\left[\frac{1}{(k+Q+q)^2-m^2}+\frac{1}{(k-Q-q)^2-m^2}\right]\right\}_{Q_0=E_Q} .
\end{eqnarray}
This procedure opens the $Q$-propagator and sets the $Q$-particle on-shell with a positive energy $Q_0=E_Q$. This leads to diagrams of the form shown in Fig.(\ref{fig4}b'), where graphs obtained by permutations of vertices as well as by the replacement $Q \rightarrow -Q$   are to be understood.   

We can similarly integrate over $q_0$ by using Cauchy's theorem. The contribution at the pole $q_0 = E_q$ coming from the equation \eqref{e56nv} is     
\begin{eqnarray}\label{e58nv}
I_{11} &=& 
  \frac{1}{2E_Q}\frac{1}{2E_q}
  \left[\frac{1}{2}+N(E_Q)\right]\left[\frac{1}{2}+N(E_q)\right]
\nonumber \\ &&
\left[\frac{1}{(k+Q+q)^2-m^2}+\frac{1}{(k-Q-q)^2-m^2}+
\frac{1}{(k+Q-q)^2-m^2}+\frac{1}{(k-Q+q)^2-m^2}
\right]_{Q_0=E_Q;\,q_0=E_q} .
\end{eqnarray}
The contribution coming from \eqref{e56nv} at the other poles may be combined with that obtained from \eqref{e57nv} at the pole $q_0 = E_q$, to give 
\begin{eqnarray}\label{e59nv}
I_{12}+I_{21} &=&
  \frac{1}{2E_Q}\frac{1}{2E_q}\left\{ \left[\frac{1}{2}+N(E_Q)\right]\left(\left[\frac{1}{2}+N(E_q-E_Q)\right]
\left[\frac{1}{(k+Q-q)^2-m^2}+\frac{1}{(k-Q+q)^2-m^2}\right]
\right. \right. \nonumber \\ &+& 
\left. \left. \left[\frac{1}{2}+N(E_Q+E_q)\right]
\left[\frac{1}{(k+Q+q)^2-m^2}+\frac{1}{(k-Q-q)^2-m^2}\right]\right)
+\left\{Q\leftrightarrow q\right\}\right\}_{Q_0=E_Q;\, q_0=E_q}.
\end{eqnarray}
Finally, the contribution obtained from the poles of the other propagators in \eqref{e57nv} is given by
\begin{eqnarray}\label{e510nv}
I_{22} &=&  \frac{1}{2E_Q}\frac{1}{2E_q}\frac{1}{2}\left\{ \left[\frac{1}{2}+N(-E_q)\right] \left[\frac{1}{2}+N(E_Q+E_q)\right]
\left[\frac{1}{(k+Q+q)^2-m^2}+\frac{1}{(k-Q-q)^2-m^2}\right]
\right.  \nonumber \\ &+& 
\left. \left[\frac{1}{2}+N(E_q)\right]\left[\frac{1}{2}+N(E_q-E_Q)\right]
\left[\frac{1}{(k+Q-q)^2-m^2}+\frac{1}{(k-Q+q)^2-m^2}\right]
+\left\{Q\leftrightarrow q\right\}\right\}_{Q_0=E_Q;\, q_0=E_q},
\end{eqnarray}
where we have symmetrized under the interchange $Q \leftrightarrow q$.

The above expressions can be simplified by using the following identities involving the statistical factors
\begin{subequations}\label{e511nv}
\be\label{e511nva}
N(E_Q-E_q)+N(E_q-E_Q) = -1
\ee
\be\label{e511nvb}
N(E_Q-E_q)-N(E_q-E_Q) = \frac{N(E_Q)+N(E_q) + 2 N(E_Q)N(E_q)}{N(E_q)-N(E_Q)}
\ee
\be\label{e511nvc}
N(E_Q+E_q) = \frac{N(E_Q)N(E_q)}{1+N(E_Q)+N(E_q)} .
\ee
\end{subequations}
In this way, the expressions given in Eqs. \eqref{e59nv}  and \eqref{e510nv}  may, respectively, be written as   
\begin{eqnarray}\label{e512nv}
I_{12}+I_{21} &=&
\frac{1}{2E_Q}\frac{1}{2E_q}\frac{1}{2}\left\{
\left[N(E_Q)+N(E_q) + 2 N(E_Q)N(E_q)\right]
\left[\frac{1}{(k+Q-q)^2-m^2}+\frac{1}{(k-Q+q)^2-m^2}\right]
                  \right. \nonumber \\ &+& 
\left. 
\left[1+N(E_Q)+N(E_q)+2N(E_Q)N(E_q)\right]
\left[\frac{1}{(k+Q+q)^2-m^2}+\frac{1}{(k-Q-q)^2-m^2}\right]
                                 \right\}_{Q_0=E_Q;\, q_0=E_q}.
\end{eqnarray}
and
\begin{eqnarray}\label{e513nv}
I_{22} &=&
-\frac{1}{2E_Q}\frac{1}{2E_q}\frac{1}{4}\left\{
\left[1+N(E_Q)+N(E_q) + 2 N(E_Q)N(E_q)\right]
\left[\frac{1}{(k+Q+q)^2-m^2}+\frac{1}{(k-Q-q)^2-m^2}\right]
                  \right. \nonumber \\ &+& 
\left. 
\left[N(E_Q)+N(E_q)+2N(E_Q)N(E_q)\right]
\left[\frac{1}{(k+Q-q)^2-m^2}+\frac{1}{(k-Q+q)^2-m^2}\right]
                                 \right\}_{Q_0=E_Q;\, q_0=E_q}.
\end{eqnarray}

The above procedure %opens as well the $q$-propagators
opens the $q$-propagator in the one-loop vertex correction 
shown in Fig.(\ref{fig4}b')
and sets the $q$ particle on-shell with a positive energy $q_0=E_q$. This leads to diagrams of the form shown in Fig. (\ref{fig4}b''), where graphs obtained by permutations and by making $q\leftrightarrow -q$ are to be understood.       

% We note here that the Eq.\eqref{e22}, which we have used in the above derivation, is valid provided the function $f$ is regular along the imaginary axis. However, as one can see in Eq. \eqref{e57nv}, the corresponding function has singularities along the imaginary $q_0$  axis at  the points               where the statistical factors $N(E_Q\mp q_0)$ have poles. To treat this case, one must accordingly deform the contour which yields additional contributions. Adding  these to the contributions given in Eqs. \eqref{e58nv}, \eqref{e512nv} and \eqref{e513nv}, it turns out that the complete result coming from Fig. (\ref{fig4}b) may  be written in the form 
We still must take into account the contributions coming from the poles of the statistical factors $N(E_Q\mp q_0)$ in the Eq. \eqref{e57nv}. These arise at the points $q_0=\pm E_Q+i 2\pi jT$, where $j$ is an integer, along two lines parallel to the imaginary $q_0$ axis. Upon using Eq. \eqref{e22}
and the identities \eqref{e511nv}, one can similarly evaluate such contributions which should be added to those given in Eqs. \eqref{e58nv}, \eqref{e512nv} and \eqref{e513nv}. Proceeding in this way, and using partial fraction decompositions like \eqref{poles1} it turns out that the complete contribution of the diagram shown in Fig. (\ref{fig4}b) may be written in the form
\begin{equation}\label{e514nv}
\frac{\lambda^2}{3!}\int\frac{d^{D-1} Q}{(2\pi)^{D-1}}\frac{1}{2 E_Q}
\int\frac{d^{D-1} q}{(2\pi)^{D-1}}\frac{1}{2 E_q}\frac{1}{2 E}
\left[F_0 + 3 N(E_Q) F_1 + 3 N(E_Q)N(E_q) F_2\right],
\end{equation}
where  $E = \sqrt{(\vec k+\vec Q+\vec q)^2+m^2}$ and 
\begin{subequations}\label{e58}
  \begin{equation}\label{e58a}
F_0 = f(E_Q,E_q,E)
\end{equation}
\begin{equation}\label{e58b} 
F_1 = f(E_Q,E_q,E) + f(-E_Q,E_q,E)
\end{equation}
\begin{equation}\label{e58c}
F_2 = f(E_Q,E_q,E) + f(-E_Q,E_q,E) + f(E_Q,-E_q,E) - f(E_Q,E_q,-E) ,
\end{equation} 
\end{subequations}
where (compare with Eq.  \eqref{e212})
\begin{equation}\label{e516nv}
f(E_Q,E_q,E) = \frac{1}{k_0+i\eta+E_Q+E_q+E}+\frac{1}{-k_0-i\eta+E_Q+E_q+E} .
\end{equation}
%We note here that the zero temperature contribution given by \eqref{e516nv} arises from a combination of $T=0$ terms coming from the $1/2$ factors of Eq. \eqref{e53} and from statistical factors like $N(E_q-E_Q)$ in the zero temperature limit.
The above form of the retarded self-energy is quite convenient, especially for the calculation of its imaginary part which is related to various physical processes. In particular, the imaginary part of $F_0$ is related to the probability that the  external field creates three scalar particles at zero temperature. This can be seen from Eq. \eqref{e516nv} as well as by cutting the propagator in Fig. (\ref{fig4}b'').
(Performing the $Q$ and $q$ integrations in Eq. \eqref{e514nv}, one obtains
in the massless case at zero temperature the result given in the Appendix \ref{appC}). 
%in the massless case 

The result \eqref{e514nv} %is consistent with unitarity and 
agrees with the one obtained in the reference \cite{Parwani:1991gq} in the imaginary time formalism,  by an elaborate evaluation of the loops appearing in Fig. (\ref{fig4}b)
%(for an alternative derivation, see \cite{Brandt:1999gf}).
An alternative derivation in the real time formulation was given in Ref. \cite{Brandt:1999gf}. In this formalism, the $2\times 2$ matrix propagator is a sum of two parts, one of which represents an on-shell thermal contribution
%due to the presence of
that involves a factor $N(|p_0|)\delta(p^2-m^2)$, where $p$ is the four-momentum of the line. Such factors naturally lead to the opening of loop diagrams at non-zero temperature. However, this property does not elucidate the fact that forward scattering tree amplitudes exist as well at zero temperature.  

Based on the present formulation, we may interpret the contributions coming from the thermal self-energy diagram (\ref{fig4}b) at two loops as follows.  $F_0$ corresponds to a forward scattering amplitude of two on-shell particles at zero temperature; % (see also Appendix \ref{appC}).
The contribution linear in the statistical factor might  be thought as a forward scattering amplitude of a on-shell thermal particle
% with a retarded self-energy correction evaluated at zero temperature. In turn, this one-loop correction may also be expressed in terms
and of a forward scattering amplitude of a single,  on-shell particle at zero temperature.  Finally,  the contribution quadratic in the  statistical factor may be interpreted
%(Fig. \ref{fig4}b''))
as the forward scattering amplitude of two on-shell thermal particles (for a related approach, see references \cite{Wong:2000hq,Kapusta:2001jw}).

\section{Discussion}\label{sec6}
Using the imaginary-time formalism, we have obtained the relation \eqref{barton1}
between general one-loop diagrams and on-shell forward scattering tree amplitudes which holds at all temperatures $T$ in $D$ space-time dimensions.
%(Such relations have been previously investigated
%\cite{Barton:1990fk,Frenkel:1992ts,Brandt:1993dk} for particular amplitudes at non-zero temperatures).
%By making suitable analytic continuations in the external energies \cite{Evans:1991ky},
%this relation becomes at zero temperature equivalent to that obtained at one loop in the real time formalism for retarded \cite{Caron-Huot:2010fvq} or time-ordered \cite{Rodrigo:2008fp} 
%boundary conditions.
We note that the right hand side of Eq. \eqref{barton1} has been obtained by performing first the $Q_0$ integral by contour integration. This has the effect of opening one propagator in the loop which leads to a tree, on-shell forward scattering amplitude.
At high temperatures, this relation provides an efficient way for
computing the retarded thermal contributions in gauge theories.

%On the other hand, when using dimensional regularization to evaluate loop integrals one implements first the $\vec Q$ integration by employing an analytic continuation in $D-1$ dimensions. In the dimensionally regularized integrals, one expects that the exchange of the order of integrations should not affect the final results. However, an explicit verification of this property turns out to be somewhat subtle since one must consistently take into account the appropriate analytic continuations as well as the $i\eta$ prescriptions.

As an application of Eq. \eqref{barton1}, we have calculated in quantum gravity
(coupled to scalar fields) the lowest orders retarded graviton amplitudes at all temperatures.
At zero temperature there occurs a subtlety due to the fact that the contribution from the arcs at infinity (see Fig. \ref{figa1}c), which arise when first performing the $Q_0$ integration,
is linearly divergent.
However, when performing first the $\vec Q$ integrations, it turns out that the remaining  $Q_0$ integration converges for $D < 0$ (see Appendix \ref{appB}),
in which case the contributions from the arcs at infinity vanish. Thus,  by using a consistent analytic continuation, we see that the neglect of the contributions from the arcs at infinity can be justified in dimensional regularization.
At high temperatures, the leading contributions of all 1PI retarded functions are proportional to $T^D$. This is required by the Ward identities which reflect the gauge invariance of the theory under coordinate transformations.

%We have applied the above method to the evaluation of the thermal self-energy at two loops.
%Based on this analysis, an extension of this result to more general 1PI Green's functions at two loop might be written in the form
%Some basic features of this calculation may be extended further. To this end, we note that in the imaginary time formalism, one can have at most one statistical factor for each loop (see Eq. \eqref{e22}). Moreover, at two loops, there are only two independent internal energies to be integrated by contour integration. This procedure leads to an opening of a single internal line for each loop, so that at two loops one would have just two cut lines which set the corresponding particles on-shell. In this way, any 1PI two-loop diagram will be opened to a tree-level graph. Based on these arguments, a suitable generalization of Eq. \eqref{barton1} (for Bosonic fields) at two-loop order may be written in the form yyyyyyyy

We have applied the above method to the evaluation, in a scalar theory, of the retarded self-energy at two loops in the imaginary time formalism.
But one can show that, by using an integral reduction procedure in dimensional regularization \cite{tHooft:1978jhc}, one can express any loop integral as a linear combination of scalar loop integrals. An example of this procedure in quantum gravity was given in section \ref{sec3}.
Based on these calculations, we point out some features which may hold for more generic one-particle irreducible diagrams at two loops. In this case, there are two independent internal energies which can be integrated out by contour integrations. This leads, by successive openings of the internal lines, to two cut lines which set the corresponding particles on-shell. In this way, a 1PI two-loop  diagram may be opened to a connected tree-level graph. We also note here that in the 
imaginary time formalism, one can have at most one statistical factor for each loop  (see Eq. \eqref{e22}).
However, a new feature occurs due to the presence, in intermediate calculations at two loops, of statistical factors like $N(E_q-E_Q)$ which yield non-vanishing contributions in the zero temperature limit.
%We also note here that in the imaginary time formalism, one can....
These add up with the usual $T=0$ contributions that might be naively expected (see Eqs. \eqref{e58nv}, \eqref{e59nv} and \eqref{e510nv}, respectively).
This combination is necessary in order to get a result consistent with unitarity, which corresponds to a forward physical tree scattering amplitude.
% In this case, such terms also give, together with the usual ones (see, for example Eq. \eqref{e58nv}), finite contributions at zero temperature.
%This COMBINATION is necessary to get a result consistent  with unitarity, WHICH CORRESPONDS TO A RETARDED PHYSICAL TREE SCATTERING AMPLITUDE.

Using the above properties, a simple generalization of the Eq. \eqref{barton1} (for bosonic fields) at two loop order may be written in the form
\begin{eqnarray}\label{e61}
\Gamma(k^1,\dots ,k^L,T) &=& \int\frac{d^{D-1} Q}{(2\pi)^{D-1}}\frac{1}{2 E_Q}
                           \int\frac{d^{D-1} q}{(2\pi)^{D-1}}\frac{1}{2 E_q}
                                             \left[
                             {\cal A}_0(k^1,\dots , k^L; Q,q)
\right. \nonumber \\ &&  \left.
                             +N(E_Q) {\cal A}_1(k^1,\dots , k^L; Q,q)
+N(E_Q)N(E_q) {\cal A}_2(k^1,\dots , k^L; Q,q)
  \right]_{Q_0=E_Q; q_0=E_q} ,
\end{eqnarray}
where ${\cal A}_0$ is related to a forward scattering amplitude of two on-shell particles $Q$ and $q$ at zero temperature. The contribution linear in the statistical factor $N(E_Q)$ describes a forward scattering amplitude of a thermal on-shell particle $Q$, and of an on-shell particle $q$ at zero temperature. The term quadratic in the statistical factors represents a forward scattering amplitude of two on-shell thermal particles.

Making in Eq. \eqref{e61} appropriate analytic continuations of the external energies \cite{Evans:1991ky},
% in Eq. \eqref{e61} and taking  suitable combinations of retarded/advanced functions \cite{Evans:1991ky}, one gets
one can verify that we get for planar 1PI retarded two and three point Green's functions at two loops, the same result as obtained in the real time formalism
%for 1PI retarded or time-ordered amplitudes
\cite{Brandt:1999gf}.
%We conjecture that the above representation of thermal Green's functions may be extended to higher-loop orders.
The representation given in Eq. \eqref{e61} also holds for more generic planar amplitudes at two loops.
%(such amplitudes give dominant contributions in the SU($N$) Yang-Mills theory, in the large $N$ limit).
The issue of non-planar amplitudes has been examined at zero temperature in Ref. \cite{Caron-Huot:2010fvq}.
However, a proper extension of the relation \eqref{e61} in the case of non-planar retarded thermal Green's functions at two loops is more involved and requires further investigation.

%In the zero temperature limit, by making appropriate analytic continuations 
%, the relation 
%\eqref{e61} becomes equivalent to that obtained at two-loops in the real time formalism for retarded
%\cite{Caron-Huot:2010fvq} or time-ordered \cite{Bierenbaum:2010cy} boundary conditions. 

\begin{acknowledgments}
{We would like to thank CNPq (Brazil) for financial support and to J. C. Taylor for a valuable correspondence.
  % D. G. C. M. acknowledges ...
%F. T. B. and  J. F.  thank CNPq (Brazil) for financial support. 
%F. T. B.,  J. F.,  S. M.-F. and G. S. S. S.  thank CNPq (Brazil) for financial support.}
%This work comes as an aftermath of an original 
%project developed with the support of FAPESP (Brazil),  grant number 2018/01073-5.
D.~G.~C.~M. would like to thank Roger Macleod for helpful discussions
}
\end{acknowledgments}

\appendix

\section{Forward scattering amplitude for the scalar self-energy}\label{appA}
We give here some details on the calculation of the retarded amplitude at $T=0$. Using Eq \eqref{2.3} and making the analytic continuation $k_0\rightarrow k_0+i\eta$, we obtain,
\begin{eqnarray}\label{A1}
\Pi^r_I(k) =  -\frac{g^2 \mu^{4-D}}{2}  \int\,\frac{{d}^{D-1} Q}{(2\pi)^{D-1}}
  \frac{1}{2 |\vec Q|} 
  \left[\frac{1}{k^2+2  (k_0|\vec Q| -\vec k\cdot\vec Q) +i\eta(k_0+|\vec Q|)}
+ \frac{1}{k^2-2  (k_0|\vec Q| -\vec k\cdot\vec Q) +i\eta(k_0-|\vec Q|)}\right].
\nonumber \\
\end{eqnarray}
When computing the imaginary parts of this amplitude, for kinematical reasons it turns out that one must have $|\vec Q|<|k_0|$. Using this condition, denoting by $u$ the cosine of the angle between $\vec k$ and $\vec Q$ and doing all other angular integrations, we then obtain
\begin{eqnarray}\label{A2}
  \Pi^r_I(k) &=&  -\frac{g^2\mu^{4-D}}{2}\frac{\pi^{(D-2)/2}}{(2\pi)^{D-1}}\frac{1}{\Gamma(D/2-1)}
\int_{-1}^{1} du (1-u^2)^{(D-4)/2} 
\nonumber \\ &&
 \int_0^\infty  d |\vec Q|   |\vec Q|^{D-3}  
  \left[\frac{1}{k^2+ 2 |\vec Q| (k_0 -|\vec k| u) - i\eta \theta(-k_0)}
+       \frac{1}{k^2- 2 |\vec Q| (k_0 -|\vec k| u) + i\eta \theta(k_0)}\right].
\end{eqnarray}
{\hbox{Let us take for example, $k^2>0$,
    in which case imaginary contributions occurs.
    Then, we can re-scale}}
\\ $|\vec Q| = x k^2/2(|k_0|-|\vec k|u)\equiv x k^2/2|\vec k|(\beta-u)$,
with $\beta=|k_0|/|\vec k|$, so we get
\begin{eqnarray}\label{A3}
  \Pi^r_I(k) &=&  -\frac{g^2 \mu^{4-D}}{2}\frac{\pi^{(D-2)/2}}{(2\pi)^{D-1}}\frac{1}{\Gamma(D/2-1)}
\frac{(k^2)^{D-1}}{(2|\vec k|)^{D-2}}
                 \int_{-1}^{1} du (1-u^2)^{(D-4)/2}(\beta-u)^{2-D} 
\nonumber \\ &&
 \int_0^\infty  d x x^{D-3}  
  \left[\frac{1}{1+x\; {\rm sgn}(k_0) - i\eta \theta(-k_0)}
+       \frac{1}{1-x\; {\rm sgn}(k_0) + i\eta \theta(k_0)}\right]
\end{eqnarray}
where ${\rm sgn}$ is the sign function. One can write the expression in the square brackets
as
\begin{eqnarray}\label{A4}
\frac{\theta(k_0)}{1+x}+\frac{\theta(-k_0)}{1-x-i\eta}+\frac{\theta(-k_0)}{1+x}+
\frac{\theta(k_0)}{1-x+i\eta}=
\frac{2 \theta(k_0)}{1-x^2+i\eta}+\frac{2 \theta(-k_0)}{1-x^2-i\eta}
\end{eqnarray}
and perform the $x$ integration %by using \cite{gradshteyn},
which leads to
\begin{eqnarray}\label{A5}
  \Pi^r_I(k) &=&  -\frac{g^2 \mu^{4-D}}{2}\frac{2^{3-D}}{(4\pi)^{D/2}} \Gamma(2-D/2)
\left[
\frac{\theta(k_0)}{(-1+i\eta)^{(D-2)/2}}+\frac{\theta(-k_0)}{(-1-i\eta)^{(D-2)/2}}
                 \right]
                 \nonumber \\ &&
              \frac{(k^2)^{D-1}}{|\vec k|^{D-2}}
                 \int_{-1}^{1} du (1-u^2)^{(D-4)/2}(\beta-u)^{2-D} .
\end{eqnarray}
Similarly, doing the $u$-integration
% with the help of \cite{gradshteyn},
we obtain
(the $x$- and $u$-integrations were done using the formulas 3.241 and 3.199 of
\cite{gradshteyn}, respectively)
\begin{eqnarray}\label{A6}
  \Pi^r_I(k) &=&  -\frac{g^2 \mu^{4-D}}{2}\frac{1}{(4\pi)^{D/2}} \Gamma(2-D/2)
\frac{\Gamma^2(D/2-1)}{\Gamma(D-2)} (k^2)^{D/2-2}
                 \left[
\frac{\theta(k_0)}{(-1+i\eta)^{(D-2)/2}}+\frac{\theta(-k_0)}{(-1-i\eta)^{D/2-2}}
                 \right].
\end{eqnarray}
Rationalizing the denominators in the square bracket and recalling that we have taken $k^2>0$,
we obtain the result
\begin{eqnarray}\label{A7}
  \Pi^r_I(k) &=&  \frac{g^2 \mu^{4-D}}{2}\frac{1}{(4\pi)^{D/2}} \Gamma(2-D/2)
\frac{\Gamma^2(D/2-1)}{\Gamma(D-2)} 
                 \left[
{\theta(k_0)}{(-i\eta-k^2)^{D/2-2}}+{\theta(-k_0)}{(i\eta-k^2)^{D/2-2}}
                 \right].
\end{eqnarray}
A similar result is found when $k^2<0$ except that in this case there are no imaginary terms. Thus, we may write the above expression in the alternative form
\begin{eqnarray}\label{A8}
  \Pi^r_I(k) &=&  \frac{g^2 \mu^{4-D}}{2}\frac{1}{(4\pi)^{D/2}} \Gamma(2-D/2)
\frac{\Gamma^2(D/2-1)}{\Gamma(D-2)} 
  \left[-i\eta\; {\rm sgn}(k_0) -k^2\right]^{D/2-2}.
\end{eqnarray}
This agrees with the result obtained by using standard Euclidean space techniques
\be\label{A9}
\Pi_R(k) =  \frac{g^2}{2}\frac{1}{(4\pi)^{D/2}} \Gamma(2-D/2)
\frac{\Gamma^2(D/2-1)}{\Gamma(D-2)}  \left(-\frac{k^2}{\mu^2}\right)^{D/2-2}
\ee
together with the appropriate analytic continuation $k_0\rightarrow k_0+i\eta$.

Similarly, by making in Eq. \eqref{A9} the analytic continuation 
$k_0\rightarrow k_0+i\eta \, {\rm sgn}(k_0)$, one recovers the result obtained in the real time formalism with time-ordered boundary conditions.

We now write Eq. \eqref{2.3} in the form (using $Q_0=|\vec Q|$ and making $\vec Q\rightarrow -\vec Q$ in the second term of \eqref{e28})
\be\label{e210}
\Pi(k) =  -\frac{g^2}{2} \mu^{4-D}\int\,\frac{{d}^{D-1} Q}{(2\pi)^{D-1}}
  \frac{1}{2 |\vec Q|} 
  \left(\frac{1}{(k_0+Q_0)^2 - (\vec k+\vec Q)^2} +
        \frac{1}{(k_0-Q_0)^2 - (\vec k+\vec Q)^2} 
\right)
\ee
and define $\vec Q^\prime \equiv -(\vec k + \vec Q)$ so that
\begin{eqnarray}\label{e211}
\Pi(k) &=&  -\frac{g^2}{2} \mu^{4-D}\int\,
\frac{{d}^{D-1} Q {d}^{D-1} Q^\prime}{(2\pi)^{D-1}}
\frac{1}{2 |\vec Q|}  \frac{1}{2 |\vec Q^\prime|} 
\delta(\vec k + \vec Q + \vec Q^\prime)
 \nonumber \\ &  &
\left(\frac{1}{k_0 + |\vec Q| + |\vec Q^\prime|}-
      \frac{1}{k_0 - |\vec Q| - |\vec Q^\prime|}+
      \frac{1}{k_0 - |\vec Q| + |\vec Q^\prime|}-
      \frac{1}{k_0 + |\vec Q| - |\vec Q^\prime|} 
\right).
\end{eqnarray}
The last two terms cancel, being anti-symmetric in $|\vec Q|$
and $|\vec Q^\prime|$ leaving us with
\begin{eqnarray}\label{e212}
\Pi(k) &=&  -\frac{g^2}{2} \mu^{4-D}\int\,
\frac{{d}^{D-1} Q }{(2\pi)^{D-1}}
\frac{1}{2 |\vec Q|}  \frac{1}{2 |\vec k + \vec Q|} 
\left(\frac{1}{k_0 + |\vec Q| + |\vec k +\vec Q|}+
      \frac{1}{-k_0 + |\vec Q| + |\vec k +\vec Q|}
\right).
\end{eqnarray}
This form of $\Pi(k)$ will be useful when discussing the sunset diagram in Sec. \ref{sec5}.

%Order of integration
\section{Order of integration in $D$ dimensions}\label{appB}
Let us consider an integral of the form
\be\label{B1}
I_D(K) = \int_{-\infty} ^\infty dQ_0 \int d^{D-1} Q \frac{Q_0^p}{(Q_0^2-Q^2+K^2)^\alpha},
\ee
where $\alpha$ and $p$ are some integers and $Q=|\vec Q|$.
Performing the angular integrations, we get
\be\label{B2}
I_D(k) = \frac{2\pi^{(D-1)/2}}{\Gamma\left(\frac{D-1}{2}\right)}
\int_{-\infty} ^\infty Q_0^p dQ_0\int_0^\infty
\frac{Q^{D-2} dQ}{(Q_0^2-Q^2+K^2)^\alpha}.
\ee
Integrating first over $Q_0$, we see that the $Q_0$ integral diverges for $p>2\alpha-1$.
On the other hand, evaluating first the $Q$ integral
with the help of the formula 3.241 of \cite{gradshteyn}
\be
\int_0^{\infty} \frac{Q^{D-2} dQ}{(Q^2-Q_0^2-K^2)^\alpha}
= \frac{1}{2\Gamma(\alpha)}
\frac{\Gamma\left(\frac{D-1}{2}\right)\Gamma\left(\alpha-\frac{D-1}{2}\right)}{[-(Q_0^2+K^2)]^{\alpha-\frac{D-1}{2}}}
\ee
which converges for $D<2\alpha+1$, we obtain
\be\label{B3}
I_D(K) = (-1)^{(D-1)/2}(\pi)^{(D-1)/2}
\frac{\Gamma\left(\alpha-\frac{D-1}{2}\right)}{\Gamma(\alpha)}
\int_{-\infty}^\infty \frac{Q_0^p dQ_0}{(Q_0^2+K^2)^{\alpha-(D-1)/2}}.
\ee
Here the $Q_0$ integration converges provided $D<2\alpha-p$. For the graviton self-energy,
one has $\alpha=2$ and $p=0,\dots ,4$. In this case, the $Q_0$ integral
becomes convergent for $D<0$.

For example, for $\alpha=2$ and $p=4$, the $Q_0$ integration yields the expression
\be\label{B4}
\int_{-\infty}^{\infty} \frac{Q_0^4 dQ_0}{(Q_0^2+K^2)^{2-(D-1)/2}}
= \frac 3 4 \frac{\Gamma(1/2) \Gamma(-D/2)}{\Gamma\left(\frac{5-D}{2}\right)} (K^2)^{D/2},
\ee
which is convergent for $D<0$. Substituting \eqref{B4} into \eqref{B3}, we obtain
\be\label{B5}
\left.I_D(K)\right|_{\alpha=2,p=4} = \frac{3i}{4}\pi^{D/2}\Gamma(-D/2)(-K^2)^{D/2}.
\ee
This result may be analytically continued to positive values of $D$, where it exhibits
poles for $D=2 n$.
The above relation agrees with that obtained by using Euclidean space techniques 
and the retarded boundary condition is selected by
the prescription $K_0 \rightarrow K_0 + i \eta$.

\section{The sunset diagram}\label{appC}
The two-loop graph shown in Fig (\ref{fig4}b) may be evaluated at zero temperature, for massless particles, as follows. Using the result given in Eq. \eqref{A9}, we obtain
the contribution
\be\label{C1}
i\frac{\lambda^2}{3!} \frac{\Gamma(2-D/2)}{(4\pi)^{D/2}} B(D/2-1,D/2-1)
\int\frac{d^D q}{(2\pi)^D}\frac{1}{q^2}\frac{1}{[-(q+k)^2]^{2-D/2}}.
\ee
Using Feynman's parametrization, combining denominators and setting $p=q+x k$, we get
\be\label{C2}
i\frac{\lambda^2}{3!} \frac{\Gamma(3-D/2)}{(4\pi)^{D/2}} B(D/2-1,D/2-1)
\int_0^1 dx x^{1-D/2}
\int\frac{d^D p}{(2\pi)^D} \frac{1}{[-p^2 - x (1-x) k^2]^{3-D/2}}.
\ee
Performing the $p$ integration in Euclidean space, we arrive at
\be\label{C3}
- \frac{\lambda^2}{3!}
\frac{\Gamma(3-D)}{(4\pi)^{D}} \frac{\Gamma^2(D/2-1)}{\Gamma(D-2)}
(-k^2)^{D-3}
\int_{0}^{1} dx x^{D/2-2} (1-x)^{D-3} .
\ee
With the help of Eq. 8.380 in \cite{gradshteyn} one can do the $x$ integration which leads to
\be\label{C4}
- \frac{\lambda^2}{3!}
\frac{\Gamma(3-D)}{(4\pi)^{D}} \frac{\Gamma^3(D/2-1)}{\Gamma(3D/2-3)}
(-k^2)^{D-3}.
\ee
Such a form may be expected on Lorentz invariance and dimensions grounds.
%We note that in even space-time dimensions, there occur poles at the zeros of the sine function. 
Making in Eq. \eqref{C4} the analytic continuation $k_0 \rightarrow  k_0 + i \eta$,
one obtains the retarded sunset self-energy at zero temperature.
Setting in the above equation $D = 4 - 2 \epsilon$, one obtains
the known result for the massless sunset self-energy function. In this case, for timelike k, the corresponding imaginary part of the retarded self-energy is given by
\be\label{C5}
\frac{\lambda^2}{3!}
\frac{i\pi}{2}\frac{k^2}{(4\pi)^4}{\rm sgn}(k_0).
\ee
This result agrees with that obtained from the imaginary part of $F_0$ in Eq. \eqref{e514nv}, after integrating over the momenta $Q$ and $q$.

%\bibliography{all_new}      
%\bibliographystyle{prsty}
\end{document}